\documentclass[draft,showkeys,showpacs,eqsecnum,nofootinbib,aps]{revtex4}
\renewcommand{\theequation}{\arabic{equation}}
\def\beq{\begin{equation}}
\def\eeq{\end{equation}}
\def\bea{\begin{eqnarray}}
\def\eea{\end{eqnarray}}
\def\nn{\nonumber}

\def\na{\nabla}
\def\pa{\partial}
\def\la{\label}

\def\A{A^{^{^{\hskip -0.12cm \footnotesize{\tt o}}}}}
\begin{document}
\title{Topological aspects of dual superconductors}
\author{Soon-Tae Hong}
\email{soonhong@ewha.ac.kr} \affiliation{Department of Science
Education, Ewha Womans University, Seoul 120-750 Korea}
\affiliation{Institute for Theoretical Physics, Uppsala
University, P.O. Box 803, S-75108 Uppsala, Sweden}
\author{Antti J. Niemi}
\email{niemi@teorfys.uu.se}
\affiliation{Institute for Theoretical Physics,
Uppsala University, P.O. Box 803, S-75108 Uppsala, Sweden}
\date{\today}
\begin{abstract}
We discuss topological aspects of two-gap superconductors with and
without Josephson coupling between gaps.  We address nontrivial
topological aspects of the dual superconductors and its
connections to Meissner effect and flux quantization. The
topological knotted string geometry is also discussed in terms of
the Hopf invariant, curvature and torsion of the strings
associated with U(1)$\times$U(1) gauge group.

\end{abstract}
\maketitle

\section{Introduction}
\setcounter{equation}{0}
\renewcommand{\theequation}{\arabic{section}.\arabic{equation}}

There have been considerable attempts to understand the condensed matter
phenomenology in terms of topological configurations inherited from knot
structures~\cite{niemi00prd,babaev01,babaev02prl,niemi03plb,neutronstars,volovikbook}.  The geometry of knotted solitons
was studied to show that the total linking numbers during the soliton
interactions are preserved~\cite{niemi00prd}, and the anomaly structure of the fermions
in a knotted soliton background was shown to be related to the inherent
chiral properties of the soliton~\cite{niemi03plb}. Moreover, the curvature and torsion of a
bosonic string in 3+1 dimensions were investigated~\cite{niemi03prd}
to be employed as Hamiltonian variables in a two dimensional
Ginzburg-Landau gauge field theory~\cite{landau50}.  Interactions of vortices were
also investigated~\cite{abri57,babaev01prb} in the Ginzburg-Landau theory.  In two and three
dimensions, cross over from weak- to strong-coupling superconductivities was studied to
figure out their thermodynamics~\cite{babaev01prb2}.  Quite recently,
the SU(2) Yang-Mills theory was studied to investigate a symmetry between electric and magnetic
variables~\cite{faddeev02} and also to discuss the two-band superconductors with interband Josephson
couplings~\cite{josep,babaev02,josep2,niemihepth}.  On the other hand, many experiments
and ab initio calculations show two-band superconductivity in MgB$_{2}$, for
instance as in Refs.~\cite{mgb2,bouquet01}.  The
photoemission spectroscopy of superconductor NbSe$_{2}$ indicates also the two-band
superconductivity associated with Fermi surface sheet-dependent superconductivity
in this multi-band system~\cite{yokoya01}.
 Aslo theoretical studies indicate
a possibility of two-gap superconductivity without intrinsic
Josephson effect in liquid metallic hydrogen,
 deuterium and hydrogen alloys under extreme pressures \cite{ashcroft00}-\cite{ma}.

In this paper we will investigate the two-gap superconductors by
exploiting the two-flavor Ginzburg-Landau theory, where we study the
magnetic flux quantization of two-gap superconductors.  We will explicitly
evaluate the London penetration depth and the Meissner and Josephson effects
to obtain the nontrivial topological aspects of the two-gap superconductors.
The knotted geometry will be also discussed in the framework of the bosonic strings.

\section{Model for two-gap superconductors}
\setcounter{equation}{0}
\renewcommand{\theequation}{\arabic{section}.\arabic{equation}}

Now, in order to describe the two-band
superconductors with the interband Josephson coupling~\cite{josep,babaev02,josep2,niemihepth}, we
start with the two-flavor Ginzburg-Landau theory whose free energy density is given by
\beq
F=\frac{1}{2m_1} \left| \left(\frac{\hbar}{i}\na+
\frac{2e}{c}\vec{A}\right) \Psi_1 \right|^2
+\frac{1}{2m_2}  \left| \left(\frac{\hbar}{i}\na-\frac{2e}{c}\vec{A}\right)
\Psi_2 \right|^2
+\frac{1}{8\pi}\vec{B}^2+V+\eta (\Psi_{1}^{*}\Psi_{2}+\Psi_{2}^{*}\Psi_{1}),
\la{free1}
\eeq
where $\Psi_{1}$  and $\Psi_{2}$ are order parameters for Cooper pairs of two different flavors,
$V$ is the potential of the form
$V(|\Psi_{1,2}|^2)=-b_\alpha|\Psi_\alpha|^2+ \frac{1}{2}c_\alpha|\Psi_\alpha|^{4}$,
$(\alpha=1,2)$~\cite{babaev01,babaev02}.  Here we introduce $\eta$ which is a characteristic of the interband
Josephson coupling strength~\cite{josep,babaev02,josep2,niemihepth}.  In the case of $\eta=0$
vanishing Josephson coupling, we can describe the liquid  metalic
hydrogen which should allow coexistent superconductivity of protonic and
electronic cooper pairs~\cite{ashcroft00,ashcroft04,bonev04,ma}.  Moreover, the interband Josephson
coupling merely changes the energy of the knot associated with the two-band superconductors.
The two condensates are then characterized
by different effective masses $m_\alpha$, coherence lengths $\xi_\alpha=\hbar/(2m_\alpha b_\alpha)^{1/2}$ and
densities $\langle|\Psi_\alpha|^2\rangle={b_\alpha}/{c_\alpha}$.

We then introduce fields $\rho$ and $z_{\alpha}$ defined as
\beq
\Psi_\alpha = (2m_\alpha)^{1/2}\rho z_\alpha
\la{psi}
\eeq
where the modulus field $\rho$ is given by condensate densities and masses,
$\rho^2=\frac{1}{2m_{1}}|\Psi_1|^2+ \frac{1}{2m_{2}}|\Psi_2|^2$, and the
$CP^{1}$ complex fields $z_\alpha$ are chosen to satisfy the geometrical constraint
\beq
z_{\alpha}^{*}z_{\alpha}=|z_1|^2+|z_2|^2=1.
\label{const1}
\eeq
In the two-gap superconductors, the gauge invariant supercurrent is given by
~\cite{babaev01}
\bea
\vec{J}&=&-\frac{e}{2m_{1}}\left[\Psi_1^*\left(\frac{\hbar}{i}\na
+\frac{2e}{c}\vec{A}\right)\Psi_1
-\Psi_1\left(\frac{\hbar}{i}\na-\frac{2e}{c}\vec{A}\right)\Psi_1^{*}\right]\nn\\
& &+\frac{e}{2m_{2}}\left[\Psi_2^*\left(\frac{\hbar}{i}\na-\frac{2e}{c}\vec{A}\right)\Psi_2
-\Psi_2\left(\frac{\hbar}{i}\na+\frac{2e}{c}\vec{A}\right)\Psi_2^{*}\right],
\la{sucurr}
\eea
which can be rewritten in terms of the fields $\rho$ and $z_{\alpha}$ as follows,
\beq
\vec{J}=-\hbar e \rho^2 \left(\vec{C}+\frac{4e}{\hbar c}\vec{A}\right),
\la{sucurr2}
\eeq
where
\beq
\vec{C}=i(\na z^{\dagger}z-z^{\dagger}\na z)=i(z_1\na z_1^{*}-z_1^{*}\na z_1-z_2\na z_2^{*} + z_2^{*}\na z_2),
\la{curr}
\eeq
with $z=(z_1, z_2^*)$.

Since the $CP^{1}$ model is equivalent to the O(3) nonlinear sigma
model (NLSM)~\cite{bowick86} at the canonical level, one can introduce the dynamical physical fields $n_{a}$ $(a=1,2,3)$ which are mappings from the space-time manifold (or the direct product of a compact two-dimensional
Riemann surface ${\sf M}^{2}$ and the time dimension $R^{1}$) to the two-sphere $S^{2}$,
namely  $n_{a}: {\sf M}^{2}\otimes R^{1}\rightarrow S^{2}$.  On the other hand, the dynamical physical fields of the $CP^{1}$ model are $z_{\alpha}$ which map the spacetime manifold ${\sf M}^{2}\otimes R^{1}$ into
$S^{3}$, namely $z_{\alpha}: {\sf M}^{2}\otimes R^{1}\rightarrow S^{3}$.  Since $S^{3}$ is homeomorphic
to SU(2) group manifold and the $CP^{1}$ model is invariant under a local U(1) gauge symmetry
\beq
z\rightarrow e^{i\xi/2}z,
\la{u1gauge}
\eeq
for arbitrary space time dependent $\xi$~\cite{witten79}, the physical configuration space
of the $CP^{1}$ model is that of the gauge orbits which form the coset
$S^{3}/S^{1}=S^{2}=CP^{1}$.  In order to associate the
physical fields of the $CP^{1}$ model with those of the O(3) NLSM, we exploit the
projection from $S^{3}$ to $S^{2}$, namely the Hopf bundle~\cite{witten79,adda79}
\beq
n_{a}=z^{\dagger}\sigma_{a}z,
\label{bundle1}
\eeq
with the Pauli matrices $\sigma_{a}$ and the $n_{a}$ fields satisfying the geometrical constraint
$n_{a}n_{a}=1$, to yield the free energy
$$
F=\hbar^2(\na\rho)^2 +\frac{1}{4}\hbar^2\rho^2(\na n_{a})^2
+\frac{1}{4e^{2}\rho^2}\vec{J}^2+\frac{1}{8\pi}\vec{B}^{2}+V+K\rho^{2}n_{1},
$$
where $K=2\eta (m_{1}m_{2})^{1/2}$.  Introducing gauge invariant vector fields $\vec{S}$
in terms of the supercurrent $\vec{J}$ in (\ref{sucurr}),
$\vec{S}=\frac{1}{\hbar e\rho^{2}}\vec{J}$, one can arrive at the free energy density of the form
$$
F=\hbar^2(\na\rho)^2
+\frac{1}{4}\hbar^2\rho^2\left[(\na n_{a})^2 +\vec{S}^2\right]
+\frac{\hbar^2 c^2}{128 \pi e^2}\left(\na\times\vec{S}
+\frac{1}{2}\epsilon_{abc}n_{a}\na n_{b}\times\na n_{c}\right)^2+V+K\rho^{2}n_{1}.
$$

\section{Meissner effects}
\setcounter{equation}{0}
\renewcommand{\theequation}{\arabic{section}.\arabic{equation}}

Now, we discuss the Meissner effect in the
two-flavor topological NLSM, where the magnetic field $\vec{B}$ is
expressed in terms of the fields $\rho$, $n_{a}$ and $\vec{S}$,
\beq \vec{B}=\na\times\vec{A}=-\frac{\hbar
c}{4e}\left(\na\times\vec{S} +\frac{1}{2}\epsilon_{abc}n_{a}\na
n_{b}\times\na n_{c}\right). \la{hfield} \eeq Combining
(\ref{sucurr2}), (\ref{hfield}) and the identity
$\na\times\vec{C}=\frac{1}{2}\epsilon_{abc}n_{a}\na n_{b}\times\na
n_{c}$, we obtain the two-gap equation in terms of the
$\rho$ and $n_{a}$ fields, \beq
\na\times\vec{J}=-\frac{4e^{2}}{c}\rho^{2}\vec{B}+\frac{2}{\rho}\na\rho\times\vec{J}
-\frac{\hbar e}{2}\rho^{2}\epsilon_{abc}n_{a}\na n_{b}\times\na
n_{c}, \la{navecbigs} \eeq which can also be rewritten in terms of
the vector fields $\vec{S}$: $\na\times\vec{S}=-\frac{4e}{\hbar
c}\vec{B}-\frac{1}{2}\epsilon_{abc}n_{a} \na n_{b}\times\na
n_{c}$.  Note that in the two-gap equation
(\ref{navecbigs}) there exists topological contribution
proportional to $\epsilon_{abc}n_{a}\na n_{b}\times\na n_{c}$
which originates from interactions of Cooper pairs of two different flavors.

Next, we consider the Meissner effect~\cite{meissner33} and the corresponding London
penetration depth in the two-gap superconductor where the Maxwell equation reads
$\na\times\vec{B}=\frac{4\pi}{c}\vec{J}$.  Here the rate of time variation is assumed to be
so slow that the displacement current can be ignored.  Combining the above Maxwell
equation with the two-gap equation (\ref{navecbigs}), we arrive at the two-gap
equations for $\vec{J}$ and $\vec{B}$
\bea
\na^{2}\vec{J}&=&\left(\frac{16\pi e^{2}}{c^{2}}+\frac{2}{\rho}\na^{2}\rho-\frac{2}{\rho^{2}}
(\na\rho)^{2}\right)\vec{J}+\frac{8e^{2}}{c}\rho\na\rho\times\vec{B}
+\frac{2}{\rho^{2}}(\na\rho\cdot\vec{J})\na\rho
+\frac{2}{\rho}\left((\na\rho\cdot\na)
\vec{J}-(\vec{J}\cdot\na)\na\rho\right)\nn\\
& &+\frac{\hbar e}{2}\rho^{2}\na\times(\epsilon_{abc}n_{a}\na n_{b}\times\na n_{c})
+\hbar e\rho\na\rho\times(\epsilon_{abc}n_{a}\na n_{b}\times\na n_{c}),\nn\\
\na^{2}\vec{B}&=&\frac{16\pi e^{2}}{c^{2}}\rho^{2}\vec{B}-\frac{8\pi}{c\rho}\na\rho\times\vec{J}
+\frac{2\pi\hbar e}{c}\rho^{2}\epsilon_{abc}n_{a}\na n_{b}\times\na n_{c}.
\la{2eom}
\eea
Note that the spatial variation of the order parameter magnitude $\na\rho$ couples the
$\vec{J}$ and $\vec{B}$ field equations.  From (\ref{2eom}), we can investigate the
two-gap Meissner effect at low temperature $T<T_{c}$ as below.

At low temperature $T<T_{c}$ where the order parameter magnitude $\rho$ vary only very slightly
over the superconductor, we obtain
$\na\times\vec{J}=-\frac{4e^{2}}{c}\rho^{2}\vec{B}
-\frac{\hbar e}{2}\rho^{2}\epsilon_{abc}n_{a}\na n_{b}\times\na n_{c}$,
so that we can arrive at the decoupled equations for the $\vec{J}$ and $\vec{B}$
\bea
\na^{2}\vec{J}&=&\frac{16\pi e^{2}}{c^{2}}\rho^{2}\vec{J}
+\frac{\hbar e}{2}\rho^{2}\na\times(\epsilon_{abc}n_{a}\na n_{b}\times\na n_{c}),\nn\\
\na^{2}\vec{B}&=&\frac{16\pi e^{2}}{c^{2}}\rho^{2}\vec{B}
+\frac{2\pi\hbar e}{c}\rho^{2}\epsilon_{abc}n_{a}\na n_{b}\times\na n_{c}.
\la{low2}
\eea
Here note that we have the topological contribution with $\epsilon_{abc}n_{a}\na
n_{b}\times\na n_{c}$.  The equation for $\vec{B}$ in (\ref{low2}) then yields the
two-gap London penetration depth
\beq
\Lambda=\left(\frac{m_{1}c^{2}}{4\pi e^{2}n_{1s}}\right)^{1/2}
\left(1+\frac{m_{1}n_{2s}}{m_{2}n_{1s}}\right)^{-1/2},
\la{pen}
\eeq
where the superfluid densities $n_{\alpha s}$ are given by
$n_{\alpha s}=2|\Psi_{\alpha}|^{2}$~\cite{ash76}.  Here, we have derived the quantity
$\Lambda$ in (\ref{pen}) in London limit when $|\Psi_{\alpha}|=$ constant and thus $\epsilon_{abc}n_{a}\na
n_{b}\times\na n_{c}=0$.  Note that the two-gap surface supercurrents screen out the
applied field to yield the two-gap Meissner effect.
Moreover the two-gap London penetration depth in (\ref{pen}) is reduced to
the single-gap London penetration depth (\ref{pen2}) below in the one-flavor limit with
$n_{2s}=0$.

Next, we consider the non-topological one-flavor limit with $n_{2s}=0$ and $\na\times\vec{C}=0$.
In this limit, (\ref{navecbigs}) and (\ref{2eom}) are reduced to the form
\bea
\na\times\vec{J}&=&-\frac{e^{2}n_{1s}}{m_{1}c}\vec{B}+\frac{1}{n_{1s}}\na n_{1s}\times\vec{J},
\la{naj2}\nn\\
\na^{2}\vec{J}&=&\left(\frac{4\pi e^{2}}{m_{1}c^{2}}n_{1s}+\frac{1}{n_{1s}}\na^{2}n_{1s}
-\frac{1}{n_{1s}^{2}}(\na n_{1s})^{2}\right)\vec{J}+\frac{e^{2}}{m_{1}c}\na n_{1s}\times\vec{B}\nn\\
& &+\frac{1}{2n_{1s}^{2}}\left((\na n_{1s}\cdot\vec{J})\na n_{1s}
+\na n_{1s}(\vec{J}\cdot\na)n_{1s}\right)
+\frac{1}{n_{1s}}\left((\na n_{1s}\cdot\na)\vec{J}-(\vec{J}\cdot\na)\na n_{1s}\right),\nn\\
\na^{2}\vec{B}&=&\frac{4\pi e^{2}}{m_{1}c^{2}}n_{1s}\vec{B}-\frac{4\pi}{cn_{1s}}\na n_{1s}\times\vec{J}.
\la{2eom2}
\eea
Note that in the more restricted low temperature limit $T<T_{c}$, we have the well-known
single-gap equation, $\na\times\vec{J}=-\frac{e^{2}n_{1s}}{m_{1}c}\vec{B}$,
$\na^{2}\vec{J}=\frac{4\pi e^{2}}{m_{1}c^{2}}n_{1s}\vec{J}$ and
$\na^{2}\vec{B}=\frac{4\pi e^{2}}{m_{1}c^{2}}n_{1s}\vec{B}$,
which yield the single-gap London penetration depth~\cite{london35}
\beq
\Lambda=\left(\frac{m_{1}c^{2}}{4\pi e^{2}n_{1s}}\right)^{1/2}
=41.9~\left(\frac{r_{s}}{a_{0}}\right)^{3/2}\left(\frac{n_{e}}{n_{1s}}\right)^{1/2}~\A,
\la{pen2}
\eeq
where $r_{s}=\left(\frac{3}{4\pi n_{e}}\right)^{1/3}$, $a_{0}$ is the Bohr
radius and $n_{e}$ is the total electron density given by $n_{e}=n_{1n}+n_{1s}$ with
the normal (superfluid) electron density $n_{1n}$ ($n_{1s}$).

Exploiting the relation in (\ref{pen2}), we can rewrite the two-gap London penetration depth (\ref{pen})
as
\beq
\Lambda=41.9~\left(\frac{r_{s}}{a_{0}}\right)^{3/2}\left(\frac{n_{e}}{n_{1s}}\right)^{1/2}
\left(1+\frac{m_{1}n_{2s}}{m_{2}n_{1s}}\right)^{-1/2}~\A.
\la{pen3}
\eeq
Note that, in the two-gap London penetration depth (\ref{pen3}), with
respect to the single-gap case we have more degrees of freedom associated with the
physical parameters $m_{2}$ and $n_{2s}$ to adjust theoretical predictions to experimental data
for the London penetration depth.

\section{Flux quantization and Josephson effects}
\setcounter{equation}{0}
\renewcommand{\theequation}{\arabic{section}.\arabic{equation}}

Now, we consider the magnetic
flux quantization of the two-gap superconductors to
discuss interspecies Cooper pair tunneling, namely the Josephson effects~\cite{josephson62}.
We consider a two-gap superconductor in the shape of a cylinder-like ring where there exists a cavity inside the inner radius.  In order to evaluate the magnetic flux inside the two-gap superconductor, we embed within the interior of the superconducting material a contour encircling the cavity.  Since at low temperature $T<T_{c}$ appreciable supercurrents can flow only near the surface of the superconductor and the order parameter magnitude $\rho$ vary only very slightly over the two-gap superconductor, integration of the
supercurrent $\vec{J}$ in (\ref{sucurr2}) over a contour vanishes
to arrive at the magnetic flux $\Phi=\oint A$ carried by vortex of the superconductor.
On the other hand, to explicitly evaluate the phase effects of the two-gap superconductor, we
parameterize the $z_{\alpha}$ fields as follows
\beq
z_{1}=|z_{1}|e^{i\phi_{1}}=e^{i\phi_{1}}\cos\frac{\theta}{2},~~~
z_{2}=|z_{2}|e^{i\phi_{2}}=e^{i\phi_{2}}\sin\frac{\theta}{2}
\label{z1z2}
\eeq
to satisfy the constraint (\ref{const1}).  After some algebra, we obtain
\beq
\vec{C}=2(|z_{1}|^{2}\na\phi_{1}-|z_{2}|^{2}\na\phi_{2}).
\la{vecsphase}
\eeq
Here note that even though there exists $\na\theta$ dependence of
$z_{\alpha}\na z_{\alpha}^{*}-z_{\alpha}^{*}\na z_{\alpha}$
$(\alpha =1,2)$ in the each flavor channels, these contributions to $\vec{C}$ cancel
each other to yield vanishing overall effects.  Since the order parameters $\Psi_{\alpha}$ are
single-valued in each flavor channels, their corresponding phases should vary
$2\pi$ times integers $p_{\alpha}$ when the ring is encircled, to yield
$\oint\na\phi_{\alpha}\cdot{\rm d}\vec{l}=2\pi p_{\alpha}$ so that we can obtain
\beq
\oint C=4\pi(|z_{1}|^{2}p_{1}-|z_{2}|^{2}p_{2}).
\la{sdl}
\eeq
Exploiting (\ref{sdl}), we arrive at
$$
|\Phi|=(|z_{1}|^{2}p_{1}-|z_{2}|^{2}p_{2})\Phi_{0},
$$
which is also written in terms of the $n_{a}$ fields to yield the fractional magnetic flux quantized with
vortex of the two-gap superconductors
\beq
|\Phi|=\frac{1}{2} \left(p_{1}-p_{2}+(p_{1}+p_{2})n_{3}\right)\Phi_{0},
\la{phipp2}
\eeq
with the fluxoid $\Phi_{0}=\frac{hc}{2e}=2.0679\times 10^{-7}$ gauss-cm$^{2}$.  Here note that
the interband Josepson coupling does not change flux quantization since its role converts circularly
symmetric vortex to a two-dimensional sin-Gordon vortex~\cite{babaev02,josep2}.
To investigate a physical meaning of the magnetic flux (\ref{phipp2}) for the
two-gap superconductor, we consider
a particular case of $p_{1}=p_{2}=1$.  In this case, we can
find the magnetic flux carried by the vortex in terms of the angle
$\theta$
$$
|\Phi|=n_{3}\Phi_{0}=\Phi_{0}\cos\theta,
$$
which shows that such a vortex can possess an arbitrary fraction of magnetic
flux quantum since $|\Phi|$ depends on the parameter $\cos\theta$ measuring the relative
densities of the two condensates in the superconductor as shown in (\ref{z1z2}).  Moreover, in the case of
$p_{1}=-p_{2}$, the magnetic flux (\ref{phipp2}) is reduced to the well-known
single-gap magnetic flux quantization,
$|\Phi|=p_{1}\Phi_{0}$, where we can readily find $\theta=0$ to yield $|z_{1}|=1$ and $|z_{2}|=0$. Note that,
exploiting the above identity (\ref{vecsphase}), $\na\times\vec{J}$ in (\ref{2eom2})
can be also rewritten in terms of the phase $\phi_{1}$ as:
$\na\times\vec{J}=-\frac{e^{2}n_{1s}}{m_{1}c}\vec{B}-\frac{\hbar e}{2m_{1}}\na n_{1s}\times\na\phi_{1}
-\frac{e^{2}}{m_{1}c}\na n_{1s}\times\vec{A}$, where we have the explicit phase dependent term.

\section{Knotted string geometry}
\setcounter{equation}{0}
\renewcommand{\theequation}{\arabic{section}.\arabic{equation}}

Now, we consider bosonic string knot geometry associated with the two-gap superconductors.  It is shown
an equivalence between the two-flavor Ginzburg-Landau theory and
a version of the O(3) NLSM introduced in
Ref.~\cite{faddeev75}.   Moreover, the model in Ref.~\cite{faddeev75} describes
topological excitations in the form of stable, finite length knotted closed
vortices~\cite{niemi97} to lead to an effective string theory~\cite{niemi02}.
This equivalence can thus imply that the two-gap superconductors
similarly support topologically nontrivial, knotted solitons.

In order to investigate the stringy features of the two-flavor
Ginzburg-Landau theory, we recall that in the Hopf bundle (\ref{bundle1}),
$n_{a}$ remains invariant under the U(1) gauge transformation (\ref{u1gauge}).
Exploiting the parameterization (\ref{z1z2}), $n_{a}$ can be rewritten in terms of the
angles $\theta$ and $\beta=\phi_{1}+\phi_{2}$,
\beq
\vec{n}=(\cos\beta\sin\theta, -\sin\beta\sin\theta, \cos\theta).
\label{vecn}
\eeq
Note that $n_{a}$ is independent of the angle $\alpha=\phi_{1}-\phi_{2}$ so that $\alpha$
can be considered as a coordinate generalization of parameter $s$ of the string coordinates
$\vec{x}(s)\in R^{3}$, which describe the knot structure involved in our two-gap
superconductor.  In fact, the knot theory in the two-gap superconductor
can be constructed in terms of a bundle of two strings.  Moreover, the U(1) gauge
transformation (\ref{u1gauge}) is related with the angle $\alpha$ in such a way that
\beq
\alpha\rightarrow \alpha+\xi,
\label{u1gauge2}
\eeq
to yield reparameterization invariance $s\rightarrow \tilde{s}(s)$.

In order to evaluate the Hopf invariant associated with the knot structure of the
two-gap superconductor, we substitute (\ref{z1z2}) into (\ref{curr}) to obtain
\beq
C=\cos\theta {\rm d}\beta+{\rm d}\alpha,
\label{ccos}
\eeq
which is also attainable from (\ref{vecsphase}).  Note that $C$ in (\ref{ccos}) transforms
under (\ref{u1gauge}) as
\beq
C\rightarrow \cos\theta {\rm d}\beta+{\rm d}(\alpha+\xi),
\label{au1gauge}
\eeq
so that $C$ can be identified as the U(1) gauge field and its exterior derivative
produces the pull-back of the area two-form on the two-sphere $S^{2}$,
$$
H={\rm d}C=\frac{1}{2}\vec{n}\cdot{\rm d}\vec{n}\wedge{\rm d}\vec{n}=\sin\theta
{\rm d}\beta \wedge{\rm d}\theta,
$$
and the corresponding dual one-form $G_{i}=\frac{1}{2}\epsilon_{ijk}H_{jk}$, which can
be rewritten in terms of the angles $\theta$ and $\beta$:
$$G=\frac{1}{2}\sin\theta{\rm d}\beta \wedge {\rm d}\theta.$$
The Hopf invariant $Q_{H}$ is then given by
$$
Q_{H}=\frac{1}{8\pi^{2}}\int H \wedge C
=\frac{1}{8\pi^{2}}\int \sin\theta{\rm d}\alpha \wedge{\rm d}\beta
\wedge{\rm d}\theta.
$$
Note that if there exists a nonvanishing Hopf invariant, the bundle of two strings
forms a knot so that the flat connection ${\rm d}\alpha$ cannot be removed
through the gauge transformation (\ref{au1gauge}).

Next, to figure out the knot structure more geometrically we employ a right-handed
orthonormal basis defined by a triplet $(\vec{n},\vec{e}_{1},\vec{e}_{2})$ where $\vec{n}$ is
given by (\ref{vecn}) and
$$
\vec{e}_{1}=(\cos\beta\cos\theta,-\sin\beta\cos\theta,-\sin\theta),~~~
\vec{e}_{2}=(\sin\beta,\cos\beta,0).
$$
Using this orthonormal basis, we define with $\vec{e}_{\pm}=\vec{e}_{2}\pm i\vec{e}_{1}$ a curvature
and a torsion:
\bea
\kappa_{i}^{\pm}&=&\frac{1}{2}e^{\pm\alpha}\vec{e}_{\pm}\cdot\pa_{i}\vec{n}
=\frac{1}{2}e^{\pm\alpha}(-\sin\theta\pa_{i}\beta\pm i\pa_{i}\theta),\nn\\
\tau_{i}&=&\frac{i}{2}\vec{e}_{-}\cdot(\pa_{i}+i\pa_{i}\alpha)\vec{e}_{+}
=\cos\theta\pa_{i}\beta-\pa_{i}\alpha.\nn
\eea
Here one can readily check that the curvature $\kappa_{i}^{\pm}$ and the torsion
$\tau_{i}$ are invariant under the U(1)$\times$U(1) gauge transformations defined
by (\ref{u1gauge}) and (\ref{u1gauge2}) and also they are not independent to yield
flatness relations between them,
$$
{\rm d}\tau+2i\kappa^{+} \wedge \kappa^{-}=0,~~~{\rm d}\kappa^{\pm}\pm i\tau \wedge \kappa^{\pm}=0.
$$
Here we emphasize that the knotted stringy structures of the two-gap superconductors are
constructed only in terms of the $CP^{1}$ complex fields $z_{\alpha}$ in the order parameters
$\Psi_{\alpha}$ in (\ref{psi}), since the modulus field $\rho$ associated with the condensate
densities does not play a central role in the geometrical arguments involved in the
topological knots of the system.

\section{Conclusions}
\setcounter{equation}{0}
\renewcommand{\theequation}{\arabic{section}.\arabic{equation}}

We have studied the current equations
in two-gap superconductor to yield
the nontrivial topological aspects and discussed its relationship to
Meissner effects. We have also discussed the knotted string geometry
in terms of the Hopf invariant, curvature and torsion of the strings
associated with U(1)$\times$U(1) gauge group.

\vskip 0.3cm
The work of AJN is supported by grant VR-2003-3466.  STH would like to thank the Institute for
Theoretical Physics at the Uppsala University for the warm hospitality during his
stay, and to acknowledge financial support in part from the Korea
Science and Engineering Foundation grant R01-2000-00015.

\end{document}